\documentclass{article}
\usepackage[utf8]{inputenc}
\usepackage{graphicx}
\usepackage{times}
\usepackage{xcolor}
\usepackage{amsmath}
\usepackage{subfigure}
\usepackage{amssymb}
\usepackage{epstopdf}
\usepackage[numbers,sort&compress]{natbib}
\usepackage{appendix}
\usepackage{ mathrsfs }
\usepackage{multirow}
\newcommand{\be}{\begin{equation}
\newcommand{\ee}{\end{equation}}}
\newcommand{\bea}{\begin{eqnarray}}
\newcommand{\eea}{\end{eqnarray}}
\newcommand{\nn}{\nonumber}

\begin{document}

\title{The treatment of conformable electromagnetic theory of Maxwell as a singular system }

\author{Eqab.M.Rabei$^{1,2}$,  Mohamed Ghaleb Al-Masaeed $^{3,4}$,  Dumitru Baleanu$^{5}$ \\$^1$Science Department, Faculty of Science, Jerash Private University\\ $^2$Physics Department, Faculty of Science, Al al-Bayt University,\\ P.O. Box 130040, Mafraq 25113, Jordan\\$^3$Ministry of Education, Jordan\\ $^4$Ministry of Education and Higher Education, Qatar\\$^5$Department of Mathematics, Cankaya University, Ankara, Turkey\\eqabrabei@gmail.com\\moh.almssaeed@gmail.com}

\maketitle

%\begin{history}
%\received{Day Month Year}
%\revised{Day Month Year}
%\end{history}

\begin{abstract}
The study explores the conformable electromagnetic field theory. The concept of the conformable delta function is introduced. Subsequently,  the conformable Maxwell's equations are derived.
\end{abstract}

\section{Introduction}
The Conformable calculus alongside fractional calculus, represents a contemporary mathematical approach utilized to solve some problems in physics and engineering \cite{khalil2014new,abdeljawad2015conformable,gheondea_three-point_2015}. Moreover, fractional models exhibit heightened attributes, including enhanced memory capabilities, making them more potent than conventional models \cite{khan2021robust,kumar2021wavelet}. Conformable calculus serves to extend the conventional integer order calculus techniques.\\ 

Equations involving fractional orders often offer more accurate models for real-world phenomena compared to these relying solely on integer orders, representing a generalization of the latter. The conformable derivative appears to fulfill all the criteria of the traditional derivative. Atangana et.al \cite{atangana2015new} have delved into the investigation of various aspects such as the conformable partial derivative, the conformable gradient of order $\alpha$, the conformable of order $\alpha$ as the scalar field, and the conformable curl. 

The realm of science and technology extensively employs conformable calculus \cite{chung2020effect,al2022wkb,al2021extension,al2021quantization,al2022extension,al-masaeed_analytical_2024}. Rabei and Horani \cite{rabei2018quantization} have addressed the treatment of conformable singular Lagrangians. They have also developed Hamilton- Jacobi partial differential equations tailored to conformable singular systems, quantifying the  Christ-Lee model through the WKB approximation.

Additionally, Rabei et.al \cite{rabei2023treatment}, have explored conformable  singular systems with second-class constraints, extending Dirac theory to encompass these systems and defining the conformable Poisson Bracket.

%%%%%%%%%%%%%%%%%%%%%%%%%%%%%%%%%%%%%%%%%%%%%%%%%%%%%%%%%%%%%%%%%%%%%%%%%%%%%%%%%%%%%%%%%%%%%%%%%%%%%%%%%%%%%%%%%%%%%%%%%%%%%%%%%%%%%%%%%%%%%%%%%%%%%%%%%%%%
\section{The conformable delta function and  conformable partial derivative}
The conformable derivative of a function $f:[0,\infty ]\to R$ of order  $0<\alpha \leq 1$ is defined  as follows
\be
\label{conformable}
D_\alpha(f)(x)=\lim_{\epsilon \to 0}\frac{f(x+\epsilon x^{1-\alpha})-f(x)}{\epsilon},
\ee
and the conformable integral is defined as \cite{khalil2014new} 
\be
(I^0_\alpha)(x)=\int_0^\infty f(x)dx_\alpha=\int_0^\infty x^{\alpha-1} f(x)dx.
\ee
The integration by parts is provided by Abdeljawad \cite{abdeljawad2015conformable} as
\be
\label{2.3}
\int_a^b f(x)D_\alpha g(x)dx_\alpha = f(x) g(x)|^b_a-\int_a^b g(x)D_\alpha f(x)dx_\alpha.
\ee
When $g(x)$ transform to $G(k)$ The conformable fractional Fourier transform is defined as \cite{abdullah2022conformable,cenesiz2015solutions} 
\be
\mathcal{F}_\alpha[ g(x)](k)=G_\alpha(k)=\int_{-\infty}^\infty g(x)e^{-2\pi i x\frac{ k^\alpha}{\alpha}} dx
\ee
Thus,
\be
\mathcal{F}_\alpha[G(k)](x)=g_\alpha(x)=\int_{-\infty}^\infty G(k) e^{2\pi i k\frac{ x^\alpha}{\alpha}} dk
\ee
We introduce  a new definition for the conformable  delta function in the  conformable calculus, where $g_\alpha(x)=\delta^c(x^\alpha)$, and $G(k)=1 $, yielding
\be
\delta^c(x^\alpha) = \int_{-\infty}^\infty e^{2\pi i k\frac{ x^\alpha}{\alpha}}  dk.
\ee
With this relation, the conformable  delta function can be represented as
\be
\delta^c(x^\alpha)=\lim_{B\to \infty }\frac{\alpha}{\pi x^\alpha} \sin{\left(2\pi  B\frac{ x^\alpha}{\alpha}\right)}
\ee
\textbf{Proof}. 
\bea
\nn
\delta^c(x^\alpha) &=& \int_{-\infty}^\infty e^{2\pi i k\frac{ x^\alpha}{\alpha}}  dk=\lim_{B\to \infty }\int_{-B}^B e^{2\pi i k\frac{ x^\alpha}{\alpha}}  dk
\\\nn&=&\lim_{B\to \infty }\frac{\alpha}{2\pi x^\alpha}\left[\frac{ e^{2\pi iB\frac{ x^\alpha}{\alpha}}- e^{-2\pi i B\frac{ x^\alpha}{\alpha}}}{i}\right]=\lim_{B\to \infty }\frac{\alpha}{\pi x^\alpha} \sin{\left(2\pi  B\frac{ x^\alpha}{\alpha}\right)}
\eea
According to the conformable fractional distributions \cite{hammad2020fractional}.  \\
Let $f_\alpha(x)=A e^{-\frac{x^{2\alpha}}{\alpha \sigma^2}}$, where $A$ is constant, while $\sigma$ is its standard deviation and $\sigma^2$ is variance of the distribution, $x>0$. For $f_\alpha(x)$ to be a conformable probability distribution
function (CPDF) with support $(0,\infty)$, we require
\be
\int_0^\infty f_\alpha(x) dx_\alpha=1.
\ee
Thus, we have
\bea
f_\alpha(x)=\frac{\sqrt{\alpha}}{\sigma \sqrt{2\pi}} e^{-\frac{x^{2\alpha}}{2\alpha \sigma^2}}
\eea
Next, the conformable  delta function exhibits the following properties\\
1- $\int_{-\infty}^\infty \delta^c(x^\alpha) dx_\alpha =\frac{1}{\alpha}.$\\
2-$\int_{-\infty}^\infty f(x^\alpha) \delta^c(x^\alpha-x^{'\alpha}) dx_\alpha =\frac{1}{\alpha} f(x^{'\alpha})$\\
Using the integration by parts formula \eqref{2.3}  we obtain\\
3-$\int_{-\infty}^\infty f(x^\alpha) D_\alpha\delta^c(x^\alpha) dx_\alpha =- \int_{-\infty}^\infty 
 D_\alpha f(x^\alpha)\delta^c(x^\alpha) dx_\alpha .$\\
Generalizing these properties into three dimensions is straightforward
\be
\delta^c(x^\alpha)\delta^c(y^\alpha)\delta^c(z^\alpha)=\frac{1}{\alpha^3}\delta(x)\delta(y)\delta(z).
\ee
\be
\int_{-\infty}^\infty \delta^c(x^\alpha) \delta^c(y^\alpha) \delta^c(z^\alpha)dx_\alpha dy_\alpha dz_\alpha =\frac{1}{\alpha^3}.
\ee
\be
\int_{-\infty}^\infty f(x^\alpha,y^\alpha,z^\alpha) \delta^c(x^\alpha)\delta^c(y^\alpha)\delta^c(z^\alpha) dx_\alpha dy_\alpha dz_\alpha =\frac{1}{\alpha^3}f(x'^\alpha,y'^\alpha,z'^\alpha)
\ee.\\
Following to Atangana et.al \cite{atangana2015new} the  conformable partial derivative for a function $g$ with $n$ variables $x_1, x_2, \cdots, x_n$ of order $\alpha$ in $x_i$ is defined as
\be
\frac{\partial^\alpha}{\partial x_i^\alpha} g(x_1, x_2, \cdots, x_n) = \lim_{\epsilon \to 0} \frac{f(x_1, x_2, \cdots, x_i + \epsilon x_i^{1-\alpha}, \cdots, x_n)-f(x_1, x_2, \cdots, x_i, x_n)}{\epsilon}.
\ee
Where $0<\alpha \leq 1$. Besides, according to the theorem given in Ref \cite{atangana2015new}, if the function $g(x,y)$ for which $\partial_x^\alpha [\partial_y^\beta g(x,y)]$ and $\partial_y^\alpha [\partial_x^\beta g(x,y)]$ exist where $0<\alpha \leq 1$, then $\partial_x^\alpha [\partial_y^\beta g(x,y)]=\partial_y^\alpha [\partial_x^\beta g(x,y)]$ and the conformable gradient of order $\alpha$ is defined as 
\be
\nabla^\alpha_{\Vec{x}} g(x,y,z)= \partial_x^\alpha g \hat{i} +  \partial_y^\alpha g \hat{j} +  \partial_z^\alpha g \hat{k}.
\ee
This can  be expressed as:
\be
\nabla^\alpha_{\Vec{x}} g=\sum_{i=1}^3 (\partial_i^\alpha g) \hat{e}_i,
\ee
where $\partial_i^\alpha g=\frac{\partial^\alpha}{\partial x_i}$ and $ \hat{e}_i$ is the unit vector in the $i$ direction.
While the conformable curl of order $\alpha$ is defined 
\be
\nabla^\alpha_{\Vec{x}} \times \Vec{G} = \sum_{i=1}^3 \left[\sum_{j=1}^3 \sum_{k=1}^3 \epsilon_{ijk} \partial_{i}^\alpha G_k \right] \hat{e}_i.
\ee
where $G$ is a vector field and $ \epsilon_{ijk}$ is the Levi-Civita symbol. Additionally,
the conformable divergence of order $\alpha$ of the vector $\Vec{F} $ is denoted as
\be
\nabla^\alpha_{\Vec{x}} \cdot \Vec{F}=\sum_{i=1}^3 \partial_i^\alpha F_i.
\ee
The element of conformable area is defined as 
\be
dS^\alpha=dx^\alpha dy^\alpha =x^{\alpha-1} y^{\alpha-1} dx dy,
\ee
and the element of conformable volume takes the form 
\be
dV^\alpha=dx^\alpha dy^\alpha dz^\alpha =x^{\alpha-1} y^{\alpha-1} z^{\alpha-1} dx dy dz.
\ee
%%%%%%%%%%%%%%%%%%%%%%%%%%%%%%%%%%%%%%%%%%%%%%%%%%%%%%%%%%%%%%%%%%%%%%%%%%%%%%%%%%%%%%%%%%%%%%%%%%%%%%%%
\section{Conformable Electromagnetic theory of Maxwell}
The conformable electromagnetic theory of Maxwell serves as an instructive example of conformable singular
systems \cite{rabei2018quantization}. The potentials $A_\mu$, where $m = 0,1,2,3$ are
treated as fundamental fields. Here $A_0$ represents the scalar potential while $A_i$, with $i = 1,2,3$ denote the vector potential components. Consequently, we propose the conformable Lagrangian for this field in the following form:
\be
\label{cl}
L=\frac{-1}{4} \int F^\alpha_{\mu \nu} F^{\alpha\mu \nu} d^3x.
\ee
Where $ F^\alpha_{\mu \nu}=\partial_\mu^\alpha A_\nu-\partial_\nu^\alpha A_\mu$, here $\partial_\mu^\alpha A_\mu = \frac{\partial^\alpha A_\mu}{x^\alpha_\mu}$
One may show that the conformable electromagnet theory is invariant under the Gauge transformation of the second kind
\be
A_\mu \to A_\mu + \partial_\mu^\alpha F.
\ee
Where $F$ is an arbitrary function and the conformable Lorentz Gauge is considered as 
\be
 \partial_\mu^\alpha A_\mu = 0.
\ee
The conformable Lagrangian eq.\eqref{cl} is singular. Nonetheless, we will employ the conformable singular theory \cite{rabei2018quantization}.\\
The momenta conjugate to the fields $A^\mu$ are calculated as 
\be
\Pi^\mu_\alpha= \frac{ \partial \mathscr{L}}{\partial A^\alpha_{\mu,0}}= F^{\alpha \mu 0},
\ee
where $\mathscr{L}=-\frac{1}{4}F_{\mu \nu}^\alpha F_{\alpha \mu \nu}$ and $A^\alpha_{\mu,0}=\partial_0^\alpha A_\mu = \frac{\partial^\alpha A_\mu}{\partial t^\alpha}$. 

Therefore, the time component of the momenta $\Pi^\mu_\alpha$ read as 
\be 
\label{3.5}
\Pi^0_\alpha = F^{\alpha 0 0}= \partial_0^\alpha A_0 
-\partial_0^\alpha A_0=0. 
\ee
As per to the conformable constrained theory \cite{rabei2018quantization}, this is regarded as a conformable primary constraint. The spatial components of the conformable momenta $\Pi^i_\alpha$ are defined as the conformable electric field 
\be 
\label{elec}
\Pi^{\alpha i }= F^{\alpha i 0}=\partial^{\alpha  i}A^0 - \partial^{\alpha 0}  A^i = E^{\alpha i}, \quad i=1,2,3. 
\ee
Thus, the vector $A^\alpha_\mu$ and its derivative 
$\partial_\mu^\alpha A^\nu$ are independent. The conformable Lagrangian density can be written as
\be
\mathscr{L}= - \frac{1}{2} \Pi^{\alpha i} \Pi^\alpha_i - \frac{1}{4} F^{\alpha i j } F^\alpha_{ij}, \quad j=1,2,3. 
\ee
where 
\bea
\mathscr{L}&=&-\frac{1}{4}  F^\alpha_{0\nu} F^{\alpha 0 \nu}-\frac{1}{4}F^\alpha_{i\nu} F^{\alpha i \nu}
\\\nn&=&-\frac{1}{4}F^\alpha_{00}F^{\alpha 0 0}-\frac{1}{4} F^\alpha_{0j} F^{\alpha 0j} - \frac{1}{4} F^\alpha_{i0} F^{\alpha i0} - \frac{1}{4} F^\alpha_{ij}  F^{\alpha i j },
\\\nn&=&-\frac{1}{2} F^\alpha_{i0}  F^{\alpha i 0 }- \frac{1}{4}F^\alpha_{ij}  F^{\alpha i j }.
\eea
Noting that $F_{00}=0$ and $F_{i0}=-F_{0i}$. The canonical conformable Hamiltonian density $\mathcal{H}_0$ calculated as
\bea
\label{26}
\mathcal{H}^\alpha_0&=&\Pi^{\alpha i} \partial_0^\alpha A_i-\mathscr{L},
\\\nn&=&\Pi^{\alpha i} \partial_0^\alpha A_i + \frac{1}{2} \Pi^{\alpha i} \Pi^\alpha_i +\frac{1}{4} F_{ij}^\alpha F^{\alpha i j}.
\eea
Utilizing  relation \eqref{elec} we can express the velocities $\partial_0^\alpha A_i$  as follows,
\be
\partial_0^\alpha A^i = -\Pi^{\alpha i}+\partial^{\alpha i} A_0,
\ee
or
\be
\label{3.9}
\partial_0^\alpha A_i = -\Pi_i^{\alpha } +\partial_i^{\alpha } A_0.
\ee
Now, we can write the  conformable Hamiltonian density \eqref{26} using this relation as
\bea
\mathcal{H}_0^\alpha &=&\Pi^{\alpha i}\Pi_i^{\alpha }+\Pi^{\alpha i}\partial_i^{\alpha } A_0+\frac{1}{2} \Pi^{\alpha i}\Pi_i^{\alpha } + \frac{1}{4} F_{ij}^\alpha F^{\alpha i j}
\\\nn&=& \frac{1}{4} F_{ij}^\alpha F^{\alpha i j}-\frac{1}{2} \Pi^{\alpha i}\Pi_i^{\alpha }+\Pi^{\alpha i}\partial_i^{\alpha } A_0.
\eea
Therefore, the canonical conformable Hamiltonian  takes the form 
\be
H_0^\alpha = \int(\frac{1}{4} F_{ij}^\alpha F^{\alpha i j}-\frac{1}{2} \Pi^{\alpha i}\Pi_i^{\alpha }+\Pi^{\alpha i}\partial_i^{\alpha } A_0)d^3x_\alpha.
\ee
Integrating the last term by parts using the definition \eqref{2.3}, we get 
\be
\int \Pi^{\alpha i}\partial_i^{\alpha } A_0 d^3x_\alpha = -\int A_0^\alpha \partial_i^{\alpha } \Pi^{\alpha i}d^3x_\alpha.
\ee
Then,  we arrived to the following  conformable  Hamiltonian $H_0^\alpha$ 
\be
\label{3.12}
H_0^\alpha = \int(\frac{1}{4} F_{ij}^\alpha F^{\alpha i j}-\frac{1}{2} \Pi^{\alpha i}\Pi_i^{\alpha }- A_0^\alpha \partial_i^{\alpha } \Pi^{\alpha i})d^3x.
\ee
The Hamiltonian does not incorporate any velocities but comprises dynamical field coordinates $A_0, A_i$, and the momenta $\Pi^\nu$ along with their spatial derivatives. 

Following Dirac's theory of constrained systems \cite{dirac_lectures_1966}, the total time derivative of the primary constraints \eqref{3.5} must be zero. i.e 
\be
\label{3.13}
\frac{d \Pi^{\alpha 0 }}{dt} = \{\Pi^{\alpha 0 },H_0^\alpha\}=0,
\ee
This condition is recognized as the consistency requirement, where $\{,\}$ denotes the Poisson Bracket (PB). The fundamental  Poisson Bracket relations are defined as 
\be
\{\Pi_\mu^\alpha,A_\nu^\alpha\}=-\delta_{\mu \nu}\delta(\Vec{x}^\alpha-\Vec{x}'^\alpha)\delta(\Vec{y}^\alpha-\Vec{y}'^\alpha)\delta(z^\alpha-z'^\alpha)=-\delta_{\mu \nu}\delta_3^\alpha(\Vec{x}^\alpha-\Vec{x}'^\alpha).
\ee
Thus, by employing these  Poisson Bracket relations and equation \eqref{3.12}, we obtain 
\bea
\nn
&& \int \{\Pi^{\alpha 0 },\frac{1}{4} F_{ij}^\alpha F^{\alpha i j}-\frac{1}{2} \Pi^{\alpha i}\Pi_i^{\alpha }- A_0^\alpha \partial_i^{\alpha } \Pi^{\alpha i}\}d^3x_\alpha
\\&=&\int \left[\frac{1}{4} \{\Pi^{\alpha 0 }, F_{ij}^\alpha F^{\alpha i j}\}
-\frac{1}{2} \{\Pi^{\alpha 0 },\Pi^{\alpha i}\Pi_i^{\alpha }\}-\{\Pi^{\alpha 0 }, A_0^\alpha \partial_i^{\alpha } \Pi^{\alpha i}\}\right]d^3x_\alpha.
\eea
Let us calculate each term
\be
\label{3.14}
\{\Pi^{\alpha 0 }, F_{ij}^\alpha F^{\alpha i j}\}= F_{ij}^\alpha \{\Pi^{\alpha 0 }, F^{\alpha i j}\}+\{\Pi^{\alpha 0 }, F_{ij}^\alpha \}F^{\alpha i j}=0,
\ee
because $A_i^\alpha$ commute with $\Pi^{\alpha 0 }$, and the second term is zero
\be
\label{3.15}
\{\Pi^{\alpha 0 },\Pi^{\alpha i}\Pi_i^{\alpha }\}=0.
\ee
Besides,
\bea
\label{3.16}
\int \{\Pi^{\alpha 0 }, A_0^\alpha \partial_i^{\alpha } \Pi^{\alpha i}\}d^3x_\alpha &=&\int \{\Pi^{\alpha 0 }, A_0^\alpha \}\partial_i^{\alpha } \Pi^{\alpha i}d^3x_\alpha,
\\\nn&=&-\int \delta_3^\alpha(\Vec{x}^\alpha-\Vec{x}'^{\alpha}) \partial_i^\alpha \Pi^{\alpha i }d^3x_\alpha.
\eea
Noting that $d^3x_\alpha=dx_\alpha dy_\alpha dz_\alpha$. Using the properties of the conformable Dirac delta and substituting equations \eqref{3.14}\eqref{3.15} and \eqref{3.16} in equation \eqref{3.13} we obtain
\be
\label{3.17}
\frac{d \Pi^{\alpha 0 }}{d t} = - \frac{\partial_i^{\alpha } \Pi^{\alpha i}}{\alpha^3}=0.
\ee
Following Dirac's theory, this is a Secondary Constraint. Again the total time derivative of this constraint should be equal to zero.
\bea
\nn
\frac{d}{dt} \partial_i^{\alpha } \Pi^{\alpha i}&=&\{\partial_i^{\alpha } \Pi^{\alpha i},H_0^\alpha\},
\\\nn&=& \int\{\partial_i^{\alpha } \Pi^{\alpha i},\frac{1}{4} F_{ij}^\alpha F^{\alpha i j}-\frac{1}{2} \Pi^{\alpha i}\Pi_i^{\alpha }- A_0^\alpha \partial_i^{\alpha } \Pi^{\alpha i}\}d^3x_\alpha,
\\&=&0.
\eea
This condition is automatically fulfilled. Consequently, there are no additional constraints, remaining; we have identified all constraints within our problem. Equation \eqref{3.5} represents the primary Constraint while equation \eqref{3.17} delineates the  Secondary constraints. Following Dirac theory, these constraints are classified as first-class constraints because the Poisson Bracket between them is equal zero.
\be
\{\Pi^{\alpha 0}, \partial_i^{\alpha } \Pi^{\alpha i}\}=0.
\ee
Following Dirac's theory, this conformable Hamiltonian is not unique, we can incorporate the conformable first-class Constraint into the Hamiltonian, in other words, we can express the total conformable  Hamiltonian as
\be
H^\alpha_T=H^\alpha_0+\lambda^\alpha \int \Pi^{\alpha 0} d^3x_\alpha,
\ee
where $\lambda^\alpha $ is the Lagrange multipliers, it is arbitrary coefficients
\be
\label{31}
H^\alpha_T= \int (\frac{1}{4} F_{ij}^\alpha F^{\alpha i j}-\frac{1}{2} \Pi^{\alpha i}\Pi_i^{\alpha }- A_0^\alpha \partial_i^{\alpha } \Pi^{\alpha i}+ \lambda^\alpha \Pi^{\alpha 0}) d^3x_\alpha.
\ee
The equations of motion in terms of the total conformable  Hamiltonian read as 
\be
\frac{d}{dt}A^\alpha_0=\{A^\alpha_0,H^\alpha_T \}=\lambda^\alpha \int \{A^\alpha_0, \Pi^{\alpha 0}\}d^3x_\alpha=\frac{\lambda^\alpha}{\alpha^3}.
\ee
This implies that the Lagrange multipliers $\lambda^\alpha $ is the time derivative of the scalar potential $A^\alpha_0$. The second part of the equations of motion is 
\be
\frac{d}{dt}A^\alpha_i=\{A^\alpha_i,H^\alpha_T \}
\ee
The Poisson Bracket between $A^\alpha_i$ and the first and last terms in relation \eqref{31} is zero. Thus, we obtain equation \eqref{3.21}
\be
\label{3.21}
\frac{d}{dt}A^\alpha_i=-\frac{1}{2}\{A^\alpha_i,\Pi^{\alpha i}\Pi_i^{\alpha } \}- A^\alpha_0\{A^\alpha_i, \partial_i^{\alpha } \Pi^{\alpha i}\}.
\ee
Let us analyze each term
\bea
\nn
-\frac{1}{2}\{A^\alpha_i,\Pi^{\alpha i}\Pi_i^{\alpha } \}&=& -\frac{1}{2}\Pi^{\alpha i}\{A^\alpha_i,\Pi_i^{\alpha } \}-\frac{1}{2}\{A^\alpha_i,\Pi^{\alpha i} \}\Pi_i^{\alpha }
\\\nn&=&-\frac{1}{2}\Pi^{\alpha i} \delta^3_\alpha(\Vec{x}-\Vec{x'})+\frac{1}{2}\Pi_i^{\alpha }\delta^3_\alpha(\Vec{x}-\Vec{x'})
\\&=&\label{3.22}\Pi_i^{\alpha }\delta_3^\alpha(\Vec{x}^\alpha-\Vec{x}'^{\alpha}).
\eea
and
\bea
\label{3.23}
- A^\alpha_0\{A^\alpha_i, \partial_i^{\alpha } \Pi^{\alpha i}\}= A^\alpha_0 \partial_i^{\alpha } \{A^\alpha_i,  \Pi^{\alpha i}\}= A^\alpha_0 \partial_i^{\alpha }\delta^\alpha(\Vec{x}-\Vec{x'}).
\eea
Substituting  \eqref{3.22} and \eqref{3.23} in \eqref{3.21}, we get
\be
\frac{d}{dt}A^\alpha_i=\int \Pi_i^{\alpha }\delta^\alpha(\Vec{x}-\Vec{x'}) d^3x_\alpha + \int A^\alpha_0 \partial_i^{\alpha }  \delta_3^\alpha(\Vec{x}^\alpha-\Vec{x}'^{\alpha}) d^3x_\alpha.
\ee
Making use of the conformable  delta function properties, we find 
\be
\label{3.24}
\frac{d}{dt}A^\alpha_i=\frac{\Pi_i^{\alpha } }{\alpha^3} + \frac{\partial_i^{\alpha } A^\alpha_0}{\alpha^3}.
\ee
Now, let us calculate the total time derivative of the momenta 
\be
\frac{d}{dt}\Pi_0^{\alpha } = \int \{\Pi_0^{\alpha },\frac{1}{4} F_{ij}^\alpha F^{\alpha i j}-\frac{1}{2} \Pi^{\alpha i}\Pi_i^{\alpha }- A_0^\alpha \partial_i^{\alpha } \Pi^{\alpha i}+ \lambda^\alpha \Pi^{\alpha 0}\}d^3x_\alpha,
\ee
all the terms are zero except the third term. Thus, 
\bea
\frac{d}{dt}\Pi_0^{\alpha } &=&  \int \{\Pi_0^{\alpha },- A_0^\alpha \partial_i^{\alpha } \Pi^{\alpha i}\}d^3x_\alpha=-\int \partial_i^{\alpha } \Pi^{\alpha i}\{\Pi_0^{\alpha },A_0^\alpha\}d^3x_\alpha\\\nn&=&-\int \partial_i^{\alpha } \Pi^{\alpha i}\delta^3_\alpha(\Vec{x}-\Vec{x'})d^3x_\alpha=-\frac{\partial_i^\alpha \Pi_i^\alpha}{\alpha^3}=0.
\eea
It is zero, due to the presence of the secondary constraint.
The equation of motion corresponding to the spatial components of momentum is 
\bea
\frac{d}{dt}\Pi^{\alpha }_i &=&  \{\Pi_i^{\alpha }, H_T^\alpha \}= \int \{\Pi_i^{\alpha }, \frac{1}{4} F_{ij}^\alpha F^{\alpha i j} \}d^3x_\alpha
\\\nn&=&\int\left[\frac{1}{4} \{\Pi_i^{\alpha }, F_{ij}^\alpha  \} F^{\alpha i j} +\frac{1}{4} F_{ij}^\alpha \{\Pi_i^{\alpha },  F^{\alpha i j} \}\right]d^3x_\alpha
\\\nn&=&\int \frac{1}{2}\{\Pi_i^{\alpha },  F^{\alpha i j} \}F_{ij}^\alpha d^3x_\alpha= \int \left[\frac{1}{2}\{\Pi_i^{\alpha }, \partial^\alpha_i A^\alpha_j - \partial^\alpha_j A^\alpha_i \}F_{ij}^\alpha \right]d^3x_\alpha 
\\\nn&=& \int \left[ \frac{1}{2}\{\Pi_i^{\alpha }, \partial^\alpha_i A^\alpha_j\}F_{ij}^\alpha - \frac{1}{2}\{\Pi_i^{\alpha }, \partial^\alpha_j A^\alpha_i \}F_{ij}^\alpha \right]d^3x_\alpha.
\eea
Interchanging $i$ and $j$ in the second term of the above equation, we obtain 
\be
\frac{d}{dt}\Pi^{\alpha }_i=\int \left[\frac{1}{2}\{\Pi^{\alpha }_i, \partial^\alpha_i A^\alpha_j\}F_{ij}^\alpha - \frac{1}{2}\{\Pi^{\alpha }_i, \partial^\alpha_i A^\alpha_j \}F_{ji}^\alpha\right]d^3x_\alpha,
\ee
but $F_{ij}^\alpha =-F_{ji}^\alpha $ by anti-symmetry, then 
\bea
\nn
\frac{d}{dt}\Pi^{\alpha }_i&=&\int \{\Pi^{\alpha}_i, \partial^\alpha_i A^\alpha_j\}F_{ij}^\alpha d^3x_\alpha=-\int F_{ij}^\alpha \partial^\alpha_i \delta^3_\alpha(x^\alpha-x'^\alpha)  d^3x_\alpha.
\eea
Using the conformable delta function properties, we get
\be
\label{55}
\frac{d}{dt}\Pi^{\alpha }_i=\frac{\partial^\alpha_i F_{ij}^\alpha}{\alpha^3}
\ee
From the   secondary constraint, we derive one of the conformable Maxwell's equations 
\be
 \label{3.27}
\frac{\partial^\alpha_i\Pi_i^{\alpha } }{\alpha^3}= \frac{\nabla^\alpha \cdot E^\alpha }{\alpha^3}= 0.
\ee
Meanwhile, equation \eqref{3.24} leads to the second conformable Maxwell's equation:
\be
\label{3.28}
\frac{d}{dt}A^\alpha_i=-\frac{E_i^{\alpha }}{\alpha^3} + \frac{\nabla^\alpha A^\alpha_0}{\alpha^3}.
\ee
Here $A^\alpha_0$ represents the conformable scalar potential. Taking the conformable curl of both sides of eq.\eqref{3.28}, we obtain
\be
\frac{d\nabla^\alpha \times A^\alpha_i }{dt} = -\nabla^\alpha \times E^\alpha + \nabla^\alpha \times \nabla^\alpha A^\alpha_0
\ee
This leads to the second conformable Maxwell's equation;
\be
\label{59}
\frac{d\Vec{B}^\alpha }{dt} +\frac{\nabla^\alpha \times E^\alpha}{\alpha^3}=0.
\ee
Once again equation  \eqref{55} leads to 
\bea
\frac{d}{dt}\Pi^{\alpha \ell }&=&-\frac{\partial^\alpha_i F_{i\ell}^\alpha}{\alpha^3} =-\frac{\partial^\alpha_i (\partial^\alpha_j A^\alpha_\ell - \partial^\alpha_\ell A^\alpha_i )}{\alpha^3}\\\nn&=&-\frac{(\nabla^\alpha \times \nabla^\alpha\times \Vec{A}^\alpha)_\ell}{\alpha^3}=-\frac{(\nabla^\alpha \times \Vec{B}^\alpha)_\ell}{\alpha^3}.
\eea
This yields the third conformable Maxwell's equations
$\Pi^{\alpha \ell }=-\Vec{E}^\alpha$ then, which can be expressed as: 
\be
\label{61}
\frac{\nabla^\alpha \times \Vec{B}^\alpha}{\alpha^3}- \frac{d \Vec{E}^\alpha}{dt}=0.
\ee
It is evident that substituting $\alpha=1$ in equations \eqref{3.27}, \eqref{59} and \eqref{61} leads to the discovery of the traditional Maxwell's equations
%%%%%%%%%%%%%%%%%%%%%%%%%%%%%%%%%%%%%%%%%%%%%%%%%%%%%%%%%%%%%%%%%%%%%%%%%%%%%%%%%%%%%%%%%%%%%%%%%%%%%%%%
\section{Conclusions}
The conformable electromagnetic theory is regarded as a conformable singular theory. It is noted that the equations of motion yield the conformable Maxwell's equation. The parameter $\alpha$ appears in the conformable Maxwell's equations and it becomes evident that when we substitute $\alpha=1$, the conformable Maxwell's equations reduce to the traditional Maxwell's equations. 
%%%%%%%%%%%%%%%%%%%%%%%%%%%%%%%%%%%%%%%%%%%%%%%%%%%%%%%%%%%%%%%%%%%%%%%%%%%%%%
\subsection*{Acknowledgement} This work has been carried out during sabbatical leave granted to the first author Eqab.M.Rabei from  Al al-Bayt University (AABU) during the academic year 2023-2024.
%%%%%%%%%%%%%%%%%%%%%%%%%%%%%%%%%%%%%%%%%%%%%%%%%%%%%%%%%%%%%%%%%%%%%%%%%%%%%%%%%%%%%%%%%%%%%%%%%%%%%%%%%%%%%%%%%%%%%%%%%%%%%%%%%%%%%%%%%%%%%%%%%%
\bibliography{ref} 
\bibliographystyle{ieeetr}
\end{document}